\documentclass[10pt,letterpaper]{article}
\usepackage{opex3}

\usepackage{textcomp}
\usepackage{amssymb}

\newcommand{\be}{\begin{equation}}
\newcommand{\ee}{\end{equation}}

\begin{document}

\title{Demonstration of suppressed phonon tunneling losses in phononic bandgap shielded membrane resonators for high-$Q$ optomechanics}

\author{Yeghishe Tsaturyan$^1$, Andreas Barg$^1$, Anders Simonsen$^1$,\\ Luis Guillermo Villanueva$^2$, Silvan Schmid$^2$, \\Albert Schliesser$^1$ and Eugene S.~Polzik$^1$}

\address{$^1$ Niels Bohr Institute, Copenhagen University, Blegdamsvej~17, 2100~Copenhagen, Denmark\\ 
{}$^2$ Department of Micro- and Nanotechnology, Technical University of Denmark, 2800 Kongens Lyngby, Denmark}

\email{albert.schliesser@nbi.dk}
\email{polzik@nbi.dk}

\begin{abstract}
  Dielectric membranes with exceptional mechanical and optical properties present one of the most promising platforms in quantum opto-mechanics. The performance of stressed silicon nitride nanomembranes as mechanical resonators notoriously depends on how their frame is clamped to the sample mount, which in practice usually necessitates delicate, and difficult-to-reproduce mounting solutions.
  Here, we demonstrate that a phononic bandgap shield integrated in the membrane's silicon frame eliminates this dependence, by suppressing dissipation through phonon tunneling.
  We dry-etch the membrane's frame so that it assumes the form of a $\mathrm{cm}$-sized bridge featuring a 1-dimensional periodic pattern, whose phononic  density of states is tailored to exhibit one, or several, full band gaps around the membrane's high-$Q$ modes in the MHz-range. 
 We quantify the effectiveness of this phononic bandgap shield by optical interferometry measuring both the suppressed transmission of vibrations, as well as the influence of frame clamping conditions on the membrane modes.
 We find suppressions up to $40~\mathrm{dB}$ and, for three different realized phononic structures, consistently observe significant suppression of the  dependence of the membrane's modes on sample clamping---\emph{if} the mode's frequency lies in the bandgap.
 As a result, we achieve membrane mode quality factors of $5\times 10^{6}$ with samples that are tightly bolted to the $8~\mathrm{K}$-cold finger of a cryostat.
$Q\times f$-products of $6\times 10^{12}~\mathrm{Hz}$ at $300~\mathrm{K}$ and $14\times 10^{12}~\mathrm{Hz}$ at $8~\mathrm{K}$ are observed, satisfying one of the main requirements for optical cooling of mechanical vibrations to their quantum ground-state.
\end{abstract}



\bibliographystyle{osajnl}
\bibliography{bandgap_v08}

\begin{thebibliography}{10}
\newcommand{\enquote}[1]{``#1''}

\bibitem{Verbridge2006}
S.~S. Verbridge, J.~M. Parpia, R.~B. Reichenbach, L.~M. Bellan, and H.~G.
  Craighead, \enquote{High quality factor resonance at room temperature with
  nanostrings under high tensile stress,} Journal of Applied Physics
  \textbf{99}, 124304 (2006).

\bibitem{Unterreithmeier2010a}
Q.~P. Unterreithmeier, T.~Faust, and J.~P. Kotthaus, \enquote{Damping of
  nanomechanical resonators,} Physical Review Letters \textbf{105}, 027205
  (2010).

\bibitem{Schmid2011}
S.~Schmid, K.~D. Jensen, K.~H. Nielsen, and A.~Boisen, \enquote{Damping
  mechanisms in high-{Q} micro and nanomechanical string resonators,} Physical
  Review B \textbf{84}, 165307 (2011).

\bibitem{Zwickl2008}
B.~M. Zwickl, W.~E. Shanks, A.~M. Jayich, C.~Yang, C.~Bleszynski~Jayich, J.~D.
  Thomson, and J.~G.~E. Harris, \enquote{High quality mechanical and optical
  properties of commercial silicon nitride membranes,} Applied Physics Letters
  \textbf{92}, 103125 (2008).

\bibitem{Wilson2009}
D.~J. Wilson, C.~A. Regal, S.~B. Papp, and H.~J. Kimble, \enquote{Cavity
  optomechanics with stoichiometric {S}i{N} films,} Physical Review Letters
  \textbf{103}, 207204 (2009).

\bibitem{Purdy2012}
T.~P. Purdy, R.~W. Peterson, P.-L. Yu, and C.~A. Regal, \enquote{Cavity
  optomechanics with {Si}$_3${N}$_4$ membranes at cryogenic temperatures,} New
  Journal of Physics \textbf{14}, 115021 (2012).

\bibitem{Thompson2007}
J.~D. Thompson, B.~M. Zwickl, A.~M. Jayich, F.~Marquardt, S.~M. Girvin, and
  J.~G.~E. Harris, \enquote{Strong dispersive coupling of a high finesse cavity
  to a micromechanical membrane,} Nature \textbf{452}, 72--75 (2008).

\bibitem{Purdy2013}
T.~P. Purdy, R.~W. Peterson, and C.~A. Regal, \enquote{Observation of radiation
  pressure shot noise on a macroscopic object,} Science \textbf{339}, 801--804
  (2013).

\bibitem{Purdy2013a}
T.~P. Purdy, P.-L. Yu, R.~W. Peterson, N.~S. Kampel, and C.~A. Regal,
  \enquote{Strong optomechanical squeezing of light,} Physical Review X
  \textbf{3}, 031012 (2013).

\bibitem{Karuza2013}
M.~Karuza, C.~Biancofiore, M.~Bawaj, C.~Molinelli, M.~Galassi, R.~Natali,
  P.~Tombesi, G.~Di~Giuseppe, and D.~Vitali, \enquote{Optomechanically induced
  transparency in a membrane-in-the-middle setup at room temperature,} Physical
  Review A \textbf{88} (2013).

\bibitem{Bagci2013}
T.~Bagci, A.~Simonsen, S.~Schmid, L.~G. Villanueva, E.~Zeuthen, J.~Appel, J.~M.
  Taylor, A.~S{\o}rensen, K.~Usami, A.~Schliesser, and E.~S. Polzik,
  \enquote{Optical detection of radio waves through a nanomechanical
  transducer,} arXiv:1307.3467  (2013).

\bibitem{Andrews2013}
R.~W. Andrews, R.~W. Peterson, T.~P. Purdy, K.~Cicak, R.~W. Simmonds, C.~A.
  Regal, and K.~W. Lehnert, \enquote{Reversible and efficient conversion
  between microwave and optical light,} arXiv:1310.5276  (2013).

\bibitem{Hammerer2009}
K.~Hammerer, M.~Aspelmeyer, E.~Polzik, and P.~Zoller, \enquote{Establishing
  {E}instein-{P}oldosky-{R}osen channels between nanomechanics and atomic
  ensembles,} Physical Review Letters \textbf{102}, 020501 (2009).

\bibitem{Hammerer2010}
K.~Hammerer, K.~Stannigel, C.~Genes, P.~Zoller, P.~Treutlein, S.~Camerer,
  D.~Hunger, and T.~W. H{\"a}nsch, \enquote{Optical lattices with
  micromechanical mirrors,} Physical Review A \textbf{82}, 021803(R) (2010).

\bibitem{Southworth2009}
D.~R. Southworth, R.~A. Barton, S.~S. Verbridge, B.~Ilic, A.~D. Fefferman,
  H.~G. Craighead, and J.~M. Parpia, \enquote{Stress and silicon nitride: A
  crack in the universal dissipation of glasses,} Phyical Review Letters
  \textbf{102}, 225503 (2009).

\bibitem{Yu2012}
P.-L. Yu, T.~P. Purdy, and C.~A. Regal, \enquote{Control of material damping in
  high-{Q} membrane microresonators,} Physical Review Letters \textbf{108},
  083603 (2012).

\bibitem{Faust2013}
T.~Faust, J.~Rieger, M.~J. Seitner, J.~P. Kotthaus, and E.~M. Weig,
  \enquote{Signatures of two-level defects in the temperature-dependent damping
  of nanomechanical silicon nitride resonators,} arXiv:1310.3671  (2013).

\bibitem{Enss2005}
C.~Enss and S.~Hunklinger, \emph{Low Temperature Physics} (Springer, 2005),
  chap.~9, pp. 283--341.

\bibitem{Wilson-Rae2008}
I.~Wilson-Rae, \enquote{Intrinsic dissipation in nanomechanical resonators due
  to phonon tunneling,} Physical Review B \textbf{77}, 245418 (2008).

\bibitem{Jockel2011}
A.~Jockel, M.~T. Rakher, M.~Korppi, S.~Camerer, D.~Hunger, M.~Mader, and
  P.~Treutlein, \enquote{Spectroscopy of mechanical dissipation in
  micro-mechanical membranes,} Applied Physics Letters \textbf{99}, 143109
  (2011).

\bibitem{Chakram2013}
S.~Chakram, Y.~S. Patil, L.~Chang, and M.~Vengalattore, \enquote{Dissipation in
  ultrahigh quality factor {S}i{N} membrane resonators,} arXiv:1311.1234
  (2013).

\bibitem{Maldovan2009}
M.~Maldovan and E.~L. Thomas, \emph{Periodic Materials and Interference
  Lithography for Photonics, Phononics and Mechanics} (Wiley-VCH, 2009),
  chap.~7, pp. 183--213.

\bibitem{Maldovan2013}
M.~Maldovan, \enquote{Sound and heat revolutions in phononics,} Nature
  \textbf{503}, 209--217 (2013).

\bibitem{MayerAlegre2011}
T.~P. Mayer~Alegre, A.~Safavi-Naeini, M.~Winger, and O.~Painter,
  \enquote{Quasi-two-dimensional optomechanical crystals with a complete
  phononic bandgap,} Optics Express \textbf{19}, 5658--5669 (2011).

\bibitem{Yu2013}
P.-L. Yu, K.~Cicak, N.~S. Kampel, Y.~Tsaturyan, T.~P. Purdy, R.~W. Simmonds,
  and C.~A. Regal, \enquote{A phononic bandgap shield for high-$q$ membrane
  microresonators,} arXiv:1312.0962  (2013).

\bibitem{Khelif2006}
A.~Khelif, B.~Aoubiza, S.~Mohammadi, A.~Adibi, and V.~Laude, \enquote{Complete
  band gaps in two-dimensional phononic crystal slabs,} Physical Review E
  \textbf{74}, 046610 (2006).

\bibitem{Ashcroft}
N.~W. Ashcroft and N.~D. Mermin, \emph{Solid State Physics} (Cengage Learning,
  1st edition, 1976).

\bibitem{Sze}
G.~S. May and S.~M. Sze, \emph{Fundamentals of Semiconductor Fabrication}
  (Wiley, 1st edition, 2003).

\bibitem{Safavi-Naeini2010}
A.~H. Safavi-Naeini and O.~Painter, \enquote{Design of optomechanical cavities
  and waveguides on a simultaneous bandgap phononic-photonic crystal slab,}
  Optics Express \textbf{18}, 14926--14943 (2010).

\end{thebibliography}

\section{Introduction}

Highly stressed nano-strings \cite{Verbridge2006, Unterreithmeier2010a, Schmid2011} and -membranes  \cite{Zwickl2008,Wilson2009, Purdy2012} of stoichiometric silicon nitride (Si$_3$N$_4$) possess mechanical resonances at MHz frequencies  with extraordinarily low dissipation, reaching $Q$-factors up to $10^7$ and $Q$-frequency-products well beyond $10^{12}$.
In combination with large zero-point motion ($\mathcal{O}(1~\mathrm{fm})$), this renders them a promising platform for experiments in quantum optomechanics.
Indeed, since the pioneering experiment  of Thompson {\it et al.} \cite{Thompson2007}, who showed how nanomembranes can be coupled efficiently to an optical mode using the so-called membrane-in-the-middle approach, they have been employed for a number of striking demonstrations of optomechanical physics.
These include strong cooling \cite{Thompson2007,Wilson2009}, observation of radiation pressure quantum backaction \cite{Purdy2013}, squeezing \cite{Purdy2013a}, \mbox{optomechanically} induced transparency \cite{Karuza2013} and the first proof-of-principle implementations of mechanically resonant electro-optical transducers \cite{Bagci2013, Andrews2013}.

These ongoing, and further  proposed audacious experiments \cite{Hammerer2009, Hammerer2010}, all require the highest possible mechanical $Q$. 
Several recent studies  \cite{Southworth2009, Yu2012, Faust2013} indicate that an internal loss mechanism, in particular coupling to two-level systems \cite{Enss2005}, sets the ultimate upper bound for the $Q$ that can be attained.
However, losses due to phonon tunneling \cite{Wilson-Rae2008} from the membrane mode into the modes of its frame often limit the $Q$ in practice  \cite{Jockel2011, Chakram2013}.
The dissipation induced by this coupling can sometimes be evaded by minimizing the contact of the frame with the sample mount (suppressing the loss of the \emph{frame} modes)  \cite{Schmid2011, Purdy2012}, but reduced stability, control and---important for cryogenic experiments---thermal anchoring render this approach less attractive. 

\begin{figure}[htb]
	\centering
	\includegraphics[width=.6\linewidth]{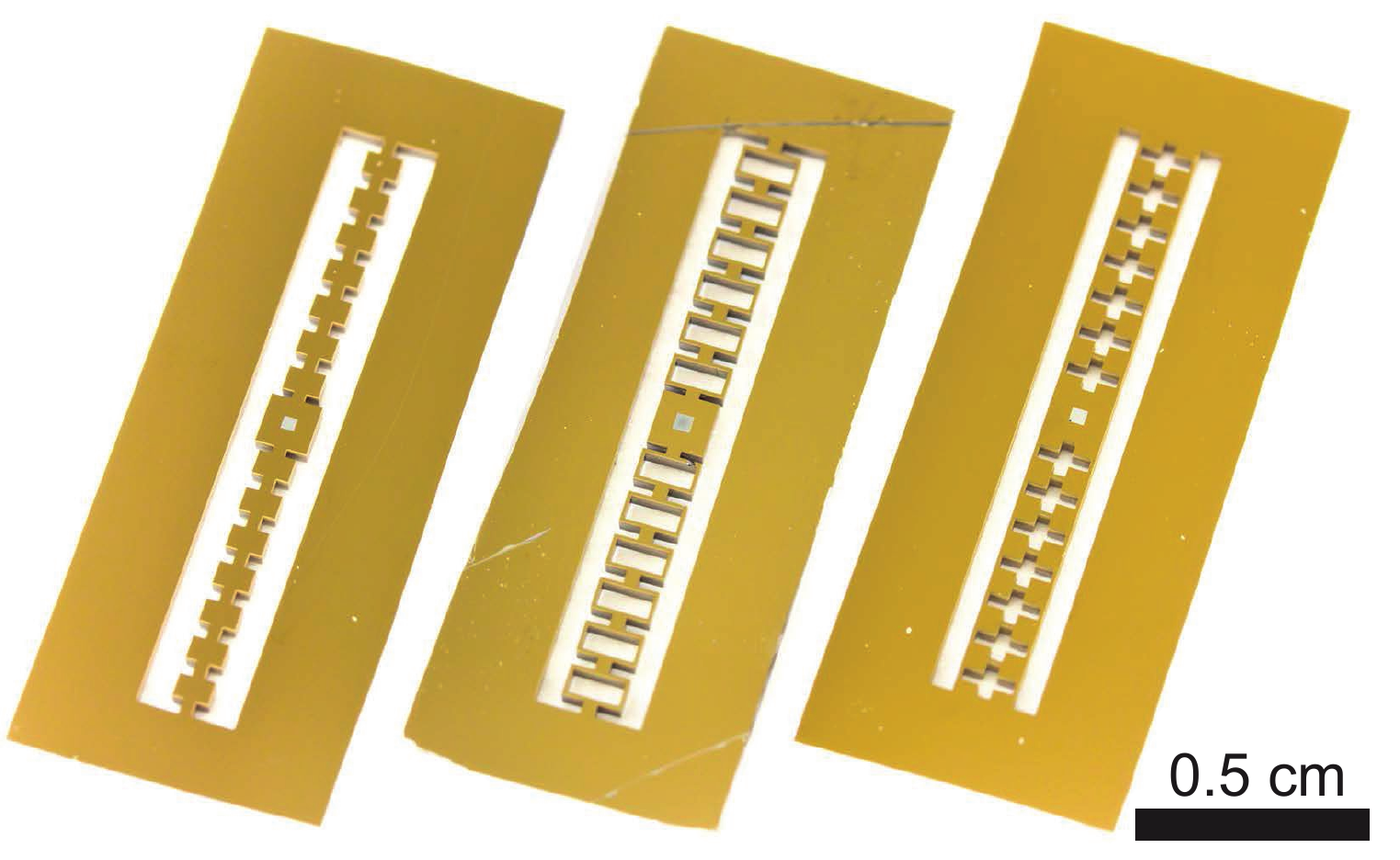}
	\caption{
	Photographs of SiN membranes held in the center of a periodically modulated silicon bridge.
	The silicon appears golden as the entire wafer has been coated with SiN, but it is etched from the backside only in the center to release the membrane.
	From left to right, three different unit cells of the periodic pattern were chosen, which we refer to as the drum structure, hollow drum structure and cross structure, respectively.
	The periodic arrangement of these cells leads to a bandgap in the phonon density of state, whose frequency span contains the membranes' mechanical modes of interest.
	 }
	\label{fig:samples}
\end{figure}

Instead, we choose here to tailor the density of phononic states in the membrane frame by patterning it with a phononic structure \cite{Maldovan2009, Maldovan2013}.
The latter are employed in science and technology at all possible  frequency and length scales, including earthquake protections, but also in surface acoustic wave devices, and most recently, optomechanical structures \cite{MayerAlegre2011, Yu2013}.
In our case, we achieve a modulation of the phononic potential landscape  by periodically removing parts of the elastic material (silicon), which gives rise to a band gap in the phonon dispersion.
The thin ($\sim 500\,\mathrm{\mu m}$) frame naturally constitutes a slab geometry, but in contrast to  previous studies \cite{Khelif2006}, we choose to adapt one-dimensional structures---essentially thin, long bridges---as a space-saving alternative with ample design flexibility, enabling particularly wide bandgaps (Fig.~\ref{fig:samples}).

\section{Simulations}
\begin{figure}[htb]
	\centering
	\includegraphics[width=\linewidth]{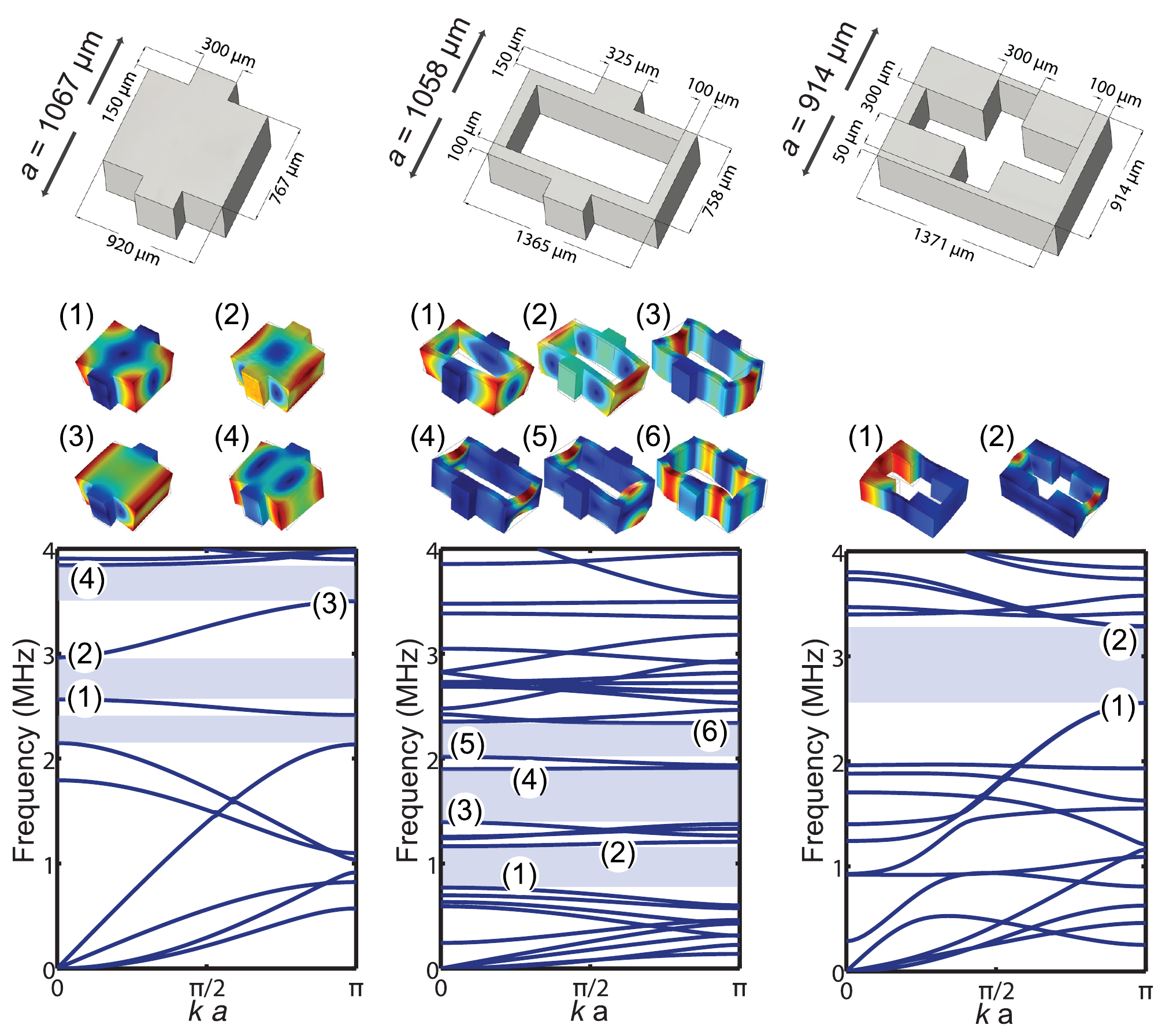}
	\caption{Unit cells (top) and band diagrams (bottom) for the drum, hollow drum and cross structures (left to right). The bandgaps relevant for our devices have been highlighted. Central row shows displacement patterns of the unit cells, as simuated numerically, for the modes at the band edges labeled in the band diagrams. The dimensions for the unit cells presented above lay the basis for the photomasks, used in the fabrication process. Also, the unit cell sizes $a$ are indicated (top).}
	\label{fig:banddiagrams}
\end{figure}

In this work we study three different quasi-one-dimensional phononic crystal structures shown in Fig. \ref{fig:samples}, each with their own characteristic features. The drum structure, mimicking a double mass-spring system with alternating masses, has three consecutive bandgaps in the frequency range of 2--4~MHz. The hollow drum structure has three bandgaps as well, however, in a lower frequency range, e.g. between 0.7 --2.4~MHz. Finally, the cross structure has one large bandgap in the range of 2.5--3.5~MHz. An overview of the band diagrams for these structure is shown in Fig.~\ref{fig:banddiagrams}.

The calculations of the band diagrams have been carried out in COMSOL through eigenfrequency analysis, employing Bloch-Floquet periodicity \cite{Ashcroft} at the boundaries of the unit cells, which are the elements of periodicity. An infinite structure with periodicity \textit{a} (e.g. the size of the unit cell in the direction of propagation, cf.~Fig.~\ref{fig:banddiagrams}) is assumed and the eigenfrequencies for different wavenumbers are calculated.
\\
\\

The structures have been optimized based on fabrication constraints and for a large relative bandgap size, $f_{c}/\Delta f$, where $f_c$ is the center-frequency of the bandgap and $\Delta f$ is the bandgap size. Importantly, we have found that wide bandgaps can only be obtained with vertical sidewalls of the holes, necessitating a physical dry etch process for patterning the silicon slab.
Full 3D simuations without restrictions with respect to the polarization of the propagating waves have been performed (cf.~Fig.~\ref{fig:banddiagrams}), thus the bandgaps obtained are complete phononic bandgaps, with no propagating modes existing in these frequency regions. 

Once the band diagrams have been calculated, the membrane mode of interest can be positioned inside a given bandgap, by adjusting the membrane size. For high-stress membrane resonators of square geometry the eigenfrequencies are given by $f_{n,m} = \sqrt{\frac{\sigma}{4\rho L^2} \left( n^2+ m^2 \right)}$, where $\sigma$ is the tensile stress ($\sim 1.14$ GPa for our devices), $\rho$ is the density of silicon nitride, $L$ is the side length of a square membrane and $(n,m)$ indicate the number of antinodes of a given membrane mode. The typical membrane size for our devices is $L~\sim 300~\mathrm{\mu m}$.

By introducing a defect in the phononic structures, thus breaking the periodicity, vibrations at certain frequencies are confined within the defect, and, reciprocally,  the defect is shielded from outside vibrations. Our devices have between 5 and 7 unit cells between the defect and the frame (see Fig. \ref{fig:samples}), the \textit{defect} being the membrane and the silicon in its immediate vicinity (i.e. the "island" in the middle of the bridge, which includes the membrane). The \textit{bridge}, consisting of the phononic structure and the defect in the middle, is thus $\sim 14~\mathrm{mm}$ long for all our devices.

Since our work is aimed at, but certainly not limited to, applications in cavity optomechanics, stable positioning of the bridge is of high importance. For the least rigid devices, the hollow drum structures, we estimate the mean square displacement of the fundamental mode of the bridge to be $\sqrt{\left\langle x^2\right\rangle} \approx 0.92$ pm at 6.2 kHz from the equipartition relation $k_\mathrm{B}T = m_{\mathrm{eff}}\Omega_\mathrm{m}^2\langle x^2\rangle$.
Here $k_\mathrm{B}$ is the Boltzmann constant, $T=300~\mathrm{K}$, and $m_{\mathrm{eff}}$ is half of the physical mass of the bridge. This estimate is in fair agreement with the measured displacement of $\sim 2~\mathrm{pm}$. The fundamental mode frequencies for the drum and cross structures are 11 kHz and 13 kHz, respectively.
\section{Fabrication}

\begin{figure}[htb]
	\centering
	\includegraphics[width=0.65\linewidth]{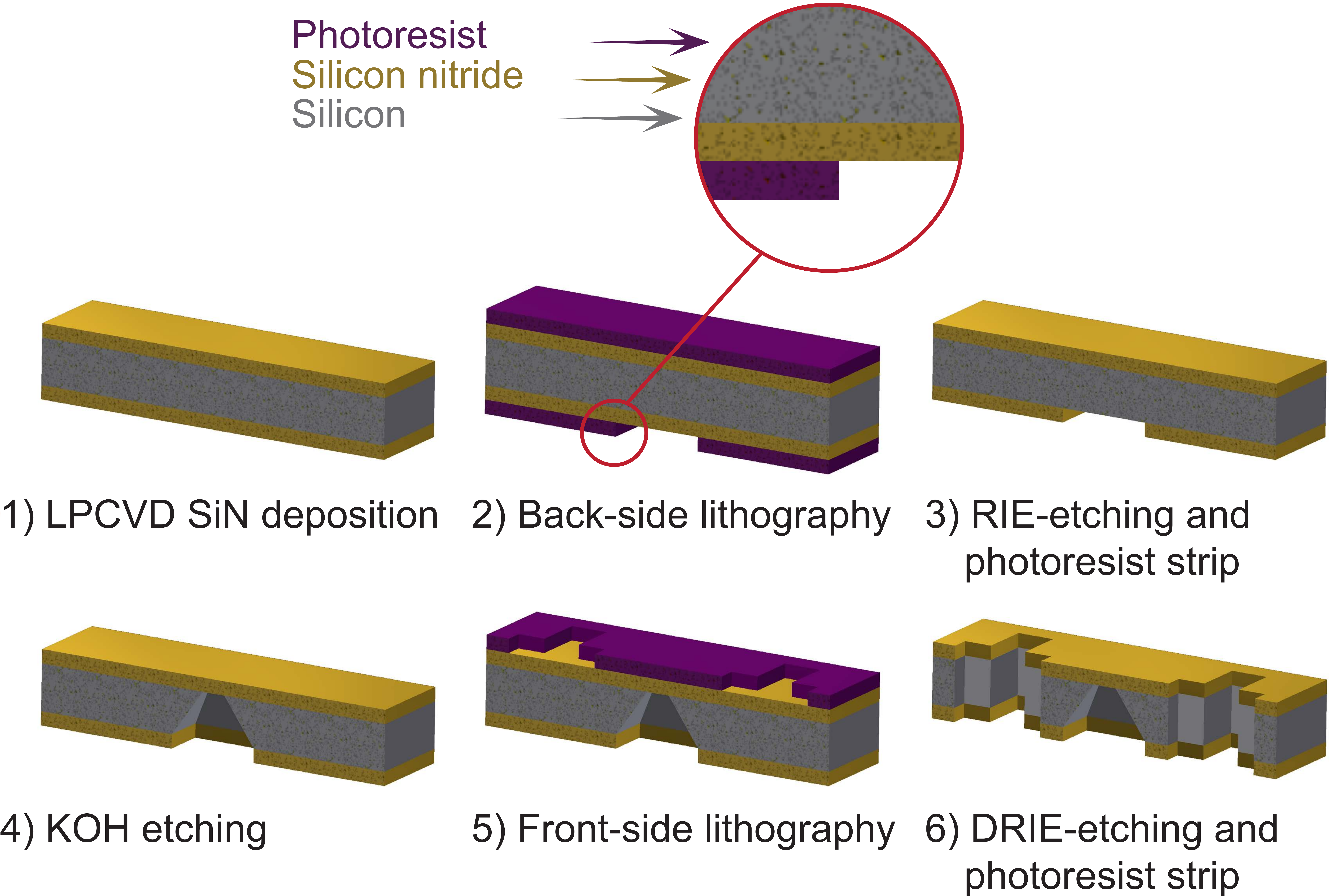}
	\caption{Illustration of the fabrication process flow. The inset in step 2 shows a cross-sectional view of the wafer after the first lithography step. Prior to photoresist removal, the wafer is exposed to chemically active plasma, removing silicon nitride and silicon ($\sim 1 \ \mu$m combined thickness) in regions with photoresist openings.}
	\label{fig:process}
\end{figure}

Our devices are fabricated in six steps (see Fig. \ref{fig:process}). We begin with low-pressure chemical vapor deposition (LPCVD) of 50 nm stoichiometric silicon nitride (SiN) on a $500~\mathrm{\mu m}$ silicon wafer. The membranes are defined by the first photolithography step on the back-side of the wafer and the developed regions are etched by reactive ion etching (RIE), opening quadratic windows in the SiN on the wafer's backside. The photoresist is removed and the wafer is stripped of its native silicon oxide by a buffered hydrofluoric acid (BHF) dip, as a preparation for the subsequent KOH etching \cite{Sze}. The square membranes are released during the KOH etch through the entire thickness of the wafer.

In the second step of photolithography the phononic structure is defined on the front-side of the wafer. The back-side of the wafer is therefore bonded to a carrier wafer using crystalbond (a water soluble mounting adhesive) and etched all the way through using deep reactive ion etching (DRIE), ensuring vertical sidewalls within $\sim 1$\textdegree. Heat removal, partly ensured by back-side gas cooling of the support wafer, is crucial during the entire etching process. Therefore, in order to prevent burning of the photoresist, the etching is done in 10 min steps. After DRIE, the wafer is placed in 80\textdegree C water for 20-30 min to dissolve the crystalbond and thus remove the carrier wafer. Finally, the photoresist is removed from the wafer using an overnight acetone bath, as well as 10-20 min of low-power oxygen plasma.
\section{Optical characterization of the phononic bandgap support}
The mechanical response of our devices is characterized using an optical interferometer. To that end, the light of a stable laser ($\lambda = 1064~\mathrm{nm}$) is reflected off a selected, $\sim 14~\mathrm{\mu m}$ large spot on the sample. The phase of the reflected light is measured by comparing it with the phase of a reference light beam in a high-bandwidth InGaAs balanced homodyne receiver (0-75 MHz). The low-frequency part of the receiver's signal is used to stabilize the reference light's phase, by displacing a mirror in the reference arm of the interferometer. This ensures measurement  of the phase quadrature of the light returned from the sample. The recorded phase quadrature is calibrated to a displacement of the probed sample area, by modulating the position of the mirror in the reference arm, at a fixed frequency, with a known amplitude. In that manner, variations of the amount of light returned from the sample, interferometric contrast etc.~are eliminated from the signal.

\begin{figure}[htb]
	\centering
	\includegraphics[width=.78\linewidth]{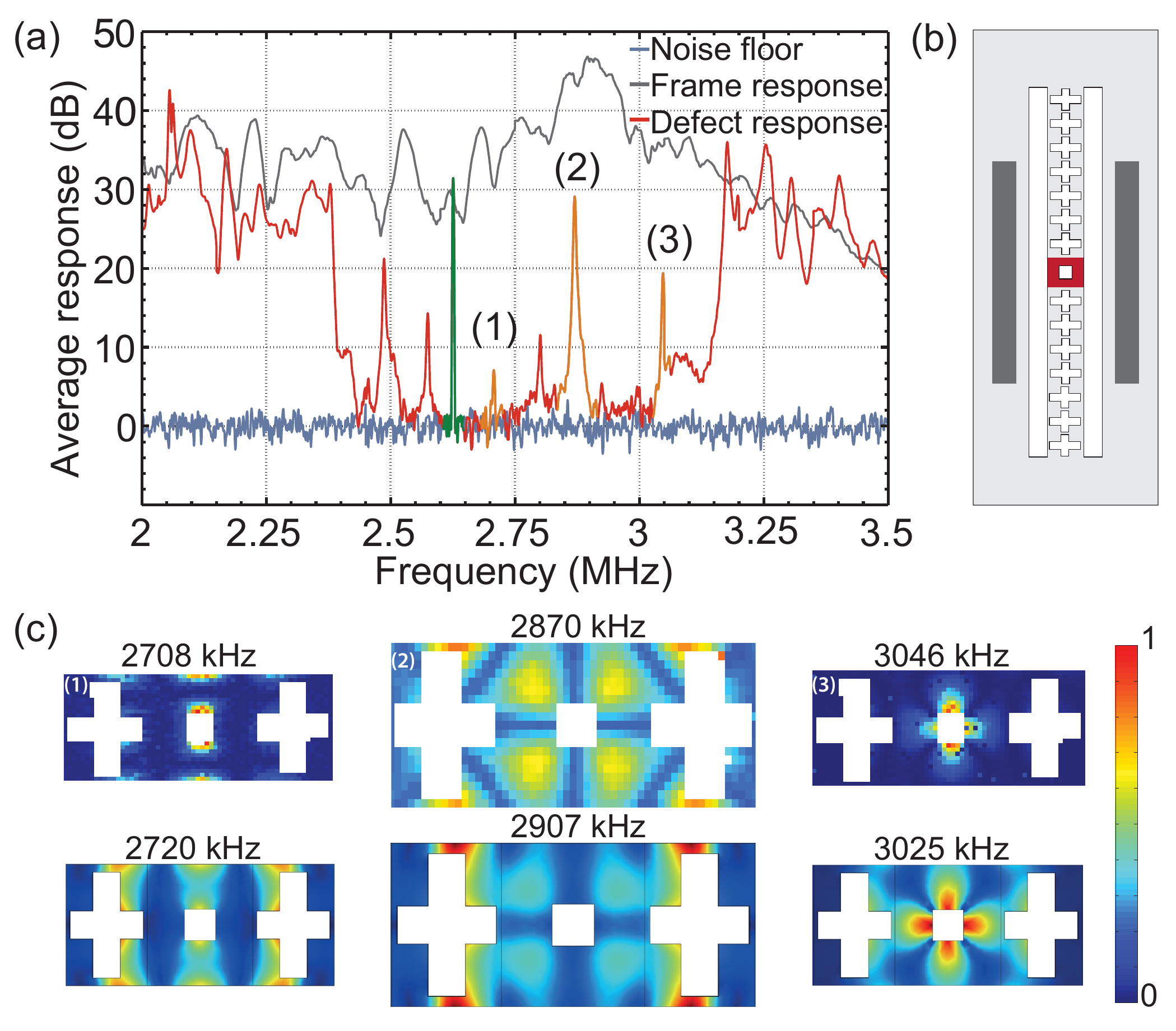}
	\caption{a) Measurement of the driven response on the defect and the frame, and the non-driven response on the defect. The data constitutes an average of 29 measurement points on the defect and the frame and has been normalized to a trace without actuation (optical shot noise, 0 dB). A calibration peak is positioned around 2.6 MHz (green). b) Illustration of a cross structure device, with the measured regions on the frame (gray) and defect (red) highlighted. c) Comparison of measured (top row) and simulated (bottom row) defect modes, indicated in the response spectrum. The colors indicates the magnitude of vertical displacement.}
	\label{fig:defectMapping}
\end{figure}

To map out the phononic bandgap we perform response measurements, i.~e., the sample is mounted to a sample holder which can be actuated (in the direction perpendicular to the plane of the sample slab) using a piezoelectric transducer. 
The sample holder is designed so that the phononic bridge can swing freely.
We measure the displacement response of the sample in several different spots using a network analyser whose output is fed to the piezo transducer, and whose input is provided by the homodyne receiver.
In particular, we compare the response measured at several spots on the frame (outside the bridge) and the defect region (in the center of the bridge, surrounding the membrane), see Fig.~\ref{fig:defectMapping}.
Measurements from several spots are averaged in order to minimize the influence of strongly localized modes of the silicon structure.
For a reference measurement, the (apparent) displacements of the frame are recorded with the actuation disabled.
With the used light powers of $\sim 3~\mathrm{mW}$, this reference trace reveals a shot noise limited sensitivity of $10~\mathrm{fm/\sqrt{Hz}}$ for the displacement measurements.

First, we consider the  response of a cross structure device. Contrary to other measurements the sample was "glued" to a spacer (e.g. bare silicon frame) using double-sided tape and subsequently attached to a piezoelectric transducer. We observe a significant drop in the average response measured on the defect in the frequency range of 2.4--3.2 MHz, reaching  up to 40 dB suppression as compared to the frame response (see Fig. \ref{fig:defectMapping}a). Comparing the observed bandgap to the predicted (see Fig. \ref{fig:banddiagrams}), we find that the bandgap sizes and center frequencies coincide to within $\sim 10\%$.

A small number of sharp resonances is observed in the bandgap. By scanning the probing spot over the sample surface, thus mapping the spatial mode profile of the response at three different resonance frequencies, we identify these as  \textit{defect modes}, i.e.~localized eigenmodes of the defect.
A comparison with a simulation of the device shows good agreement between the measured and simulated defect modes, with the resonance frequencies differing by a few tens of kilohertz (see Fig. \ref{fig:defectMapping}c). 
Note that cooling of the membrane modes can shift their frequencies by several tens of kilohertz. Therefore, being able to predict the defect modes is important for measurements at cryogenic temperatures, in order to avoid steering a membrane mode into potentially detrimental defect modes.
\section{Effect of phononic bandgap support on membrane modes}
Next, we study the effect of the phononic bandgaps of two  shield designs (cross and hollow drum structure, see Fig. \ref{fig:banddiagrams}) on the mechanical quality factors of the membranes. 
More specifically, we investigate the suppression of membrane losses due to clamping of the frame. 
For these measurements, performed at room temperature and low pressure ($\lesssim~3\times 10^{-6}~\mathrm{mbar}$), we design a small aluminum sample holder, with a removable cover, that allows tight clamping of the samples frame---while we make sure that the bridge can swing freely.
For comparison, we also measure the membrane mode's $Q$ when the sample is free-standing, i.e.~resting on the sample mount under its own gravitational weight only. 
These two configurations will be referred to as \textit{clamped} and \textit{unclamped}.

\begin{figure}[htb]
	\centering
	\includegraphics[width=1\linewidth]{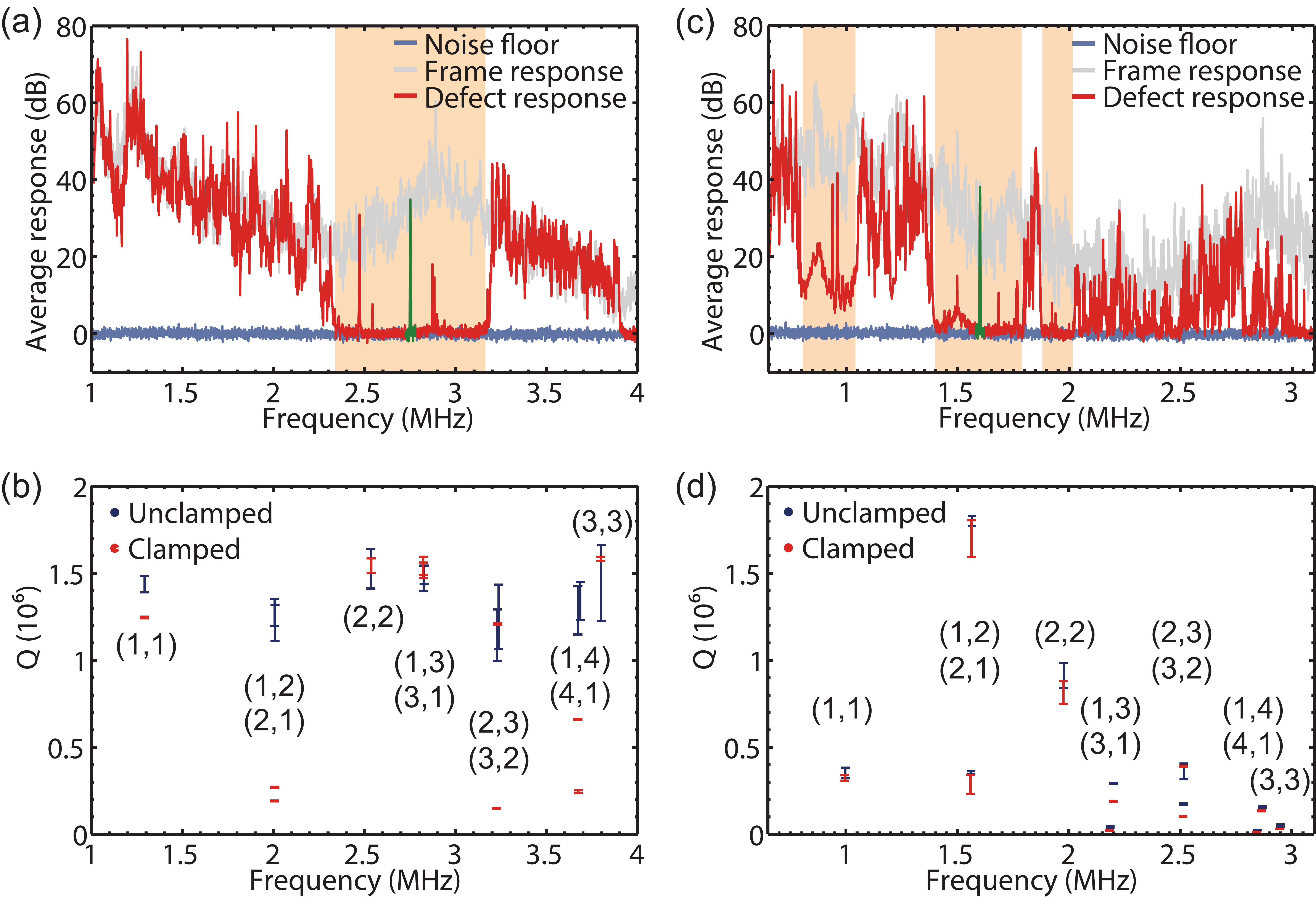}
	\caption{The response measurements of a cross structure (a-b) and a hollow drum structure device (c-d), accompanied with the mechanical $Q$-factors. The error bars for the quality factors are the standard deviation of all measured ringdowns for a given mode (typically 20 ringdown measurements for each mode). The response data is an average of 8 measurement points on the defect and the frame. The regions assumed to be the phononic bandgaps are shown in orange, while the calibration peak is indicated in green. The labels in (b) and (d) indicate the membrane modes. For the two devices the membrane sizes are (b) $L\sim 290~\mathrm{\mu m}$ and (d) $L\sim 370~\mathrm{\mu m}$.}
	\label{fig:bandgapsQs}
\end{figure}

Due to the high quality factor of the membrane modes, their $Q$ is determined using ringdown measurements.
The membrane motion is excited by a actuating the piezoelectric transducer below the sample holder.
When the optically measured displacement amplitude reaches a value close to its maximum, the actuation is switched off and the ringdown of the oscillations is recorded.
We  use  small incident light powers on the sample, typically below $200~\mathrm{\mu W}$. 
A small number of high-$Q$ modes inside the phononic bandgap required a more direct actuation, which was realised using optical forces.
The incident light power was increased to 2~mW, and modulated in amplitude, thereby exciting the modes optically. 
Again, the actuation is switched off and the ringdown recorded.
We have verified that the increase in light power doesn't affect the quality factors of the membrane modes.
For each membrane, we have measured 11 mechanical modes, ranging from the fundamental to the (3,3) mode. The results of these measurements are shown in  Fig. \ref{fig:bandgapsQs}, along with response measurements of the bridge as described previously.
\\

Both structures show suppression up to 40~dB within the phononic bandgaps. 
For the cross structure device we find good agreement with the predicted bandgap, while for the hollow drum structure only the first two bandgaps overlap with the simulation (see Fig. \ref{fig:banddiagrams}). 
In fact, all bandgaps are shifted down in frequency, compared to the simulated band diagrams. 
It is well known that by reducing the width of the connector pieces of a unit cell, the low frequency band can be pushed further down \cite{Safavi-Naeini2010}. In case of the cross structure, the connector pieces corresponds to the $50~\mathrm{\mu m}$ and $100~\mathrm{\mu m}$ thin connections perpendicular to and in the direction of propagation. 
By examining the devices under a microscope we indeed find that these connections can be up to $\sim 15\%$ thinner than predicted.
For the cross structure we find that the (2,2), as well as the (1,3) and (3,1) modes, are within the bandgap, while for the hollow drum structure the first four mechanical modes are distributed among three bandgaps. In both cases we observe that modes outside the bandgap have predominantly lower quality factors clamped versus unclamped, as one would have expected. As shown in Fig. \ref{fig:bandgapsQs}b, a majority of the membrane modes outside the bandgap are significantly affected by the clamping of the frame. In-bandgap modes, however, are virtually unaffected by the clamping of the substrate, independent of the membrane mode. These initial results already suggest that the bandgap does in fact have an effect on phonon tunneling losses of the membrane modes.

\begin{figure}[htb]
	\centering
	\includegraphics[width=.47\linewidth]{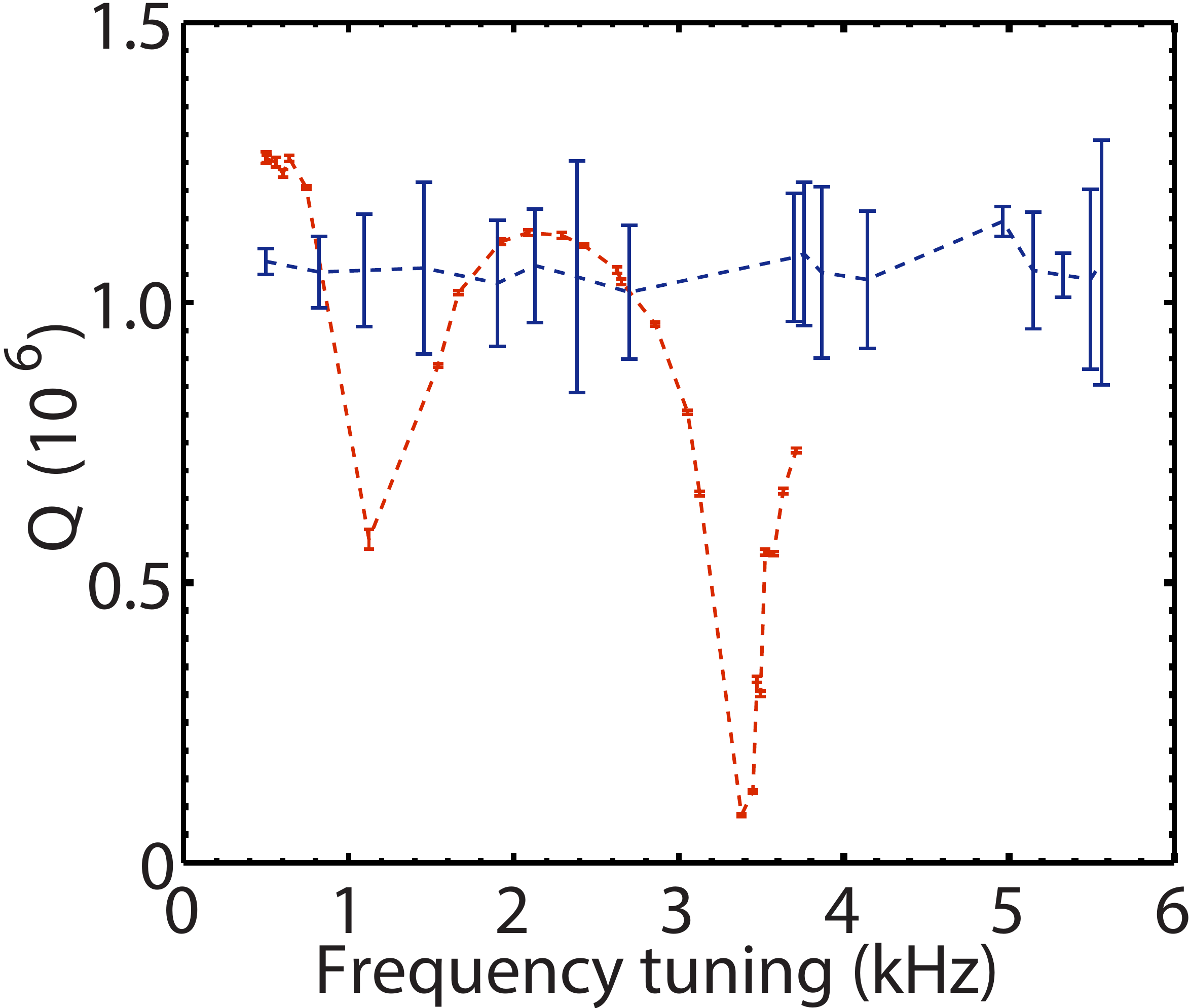}
	\caption{Frequency dependence of the $Q$-factor for a (2,2) inside (blue) and outside (red) a phononic bandgap. The change in frequency is induced by a temperature sweep from $303~\mathrm{K}$ to $319~\mathrm{K}$ (in-bandgap mode) and $326~\mathrm{K}$ (mode outside the bandgap). The starting frequencies for the in-bandgap mode is $f_0 = 2.811~\mathrm{MHz}$ and $f_0 = 3.022~\mathrm{MHz}$ for the mode outside the bandgap.}
	\label{fig:temperature}
\end{figure}

An independent indication of the successful decoupling of the membrane from the environment is found in a measurement similar to the one reported by J\"{o}ckel \textit{et al.} \cite{Jockel2011}. More specifically, we consider the frequency dependence of the quality factors, by heating the sample with an electric heater in the sample holder. After each temperature change, we wait 5--10 minutes, in order for the sample to reach thermal equilibrium, as well as the pressure to stabilize. We measure the (2,2) modes of two cross structure devices, with the membrane mode inside and outside the bandgap, respectively. As shown in Fig. \ref{fig:temperature}, the mode outside the bandgap exhibits large dips within a few kilohertz, while the quality factor of the in-bandgap mode is constant. The narrow "resonances" can be explained by the fact that the whole sample is being heated, resulting in the membrane modes moving up in frequency and the frame modes being shifted to lower frequencies \cite{Jockel2011}. By tuning the membrane mode into a frame mode the quality factor drops significantly. The density of the frame modes (sampled by the membrane in the defect) is strongly suppressed by a phononic bandgap support, therefore such resonant dips are not observed in this case.
\\

\begin{figure}[htb]
	\centering
	\includegraphics[width=\linewidth]{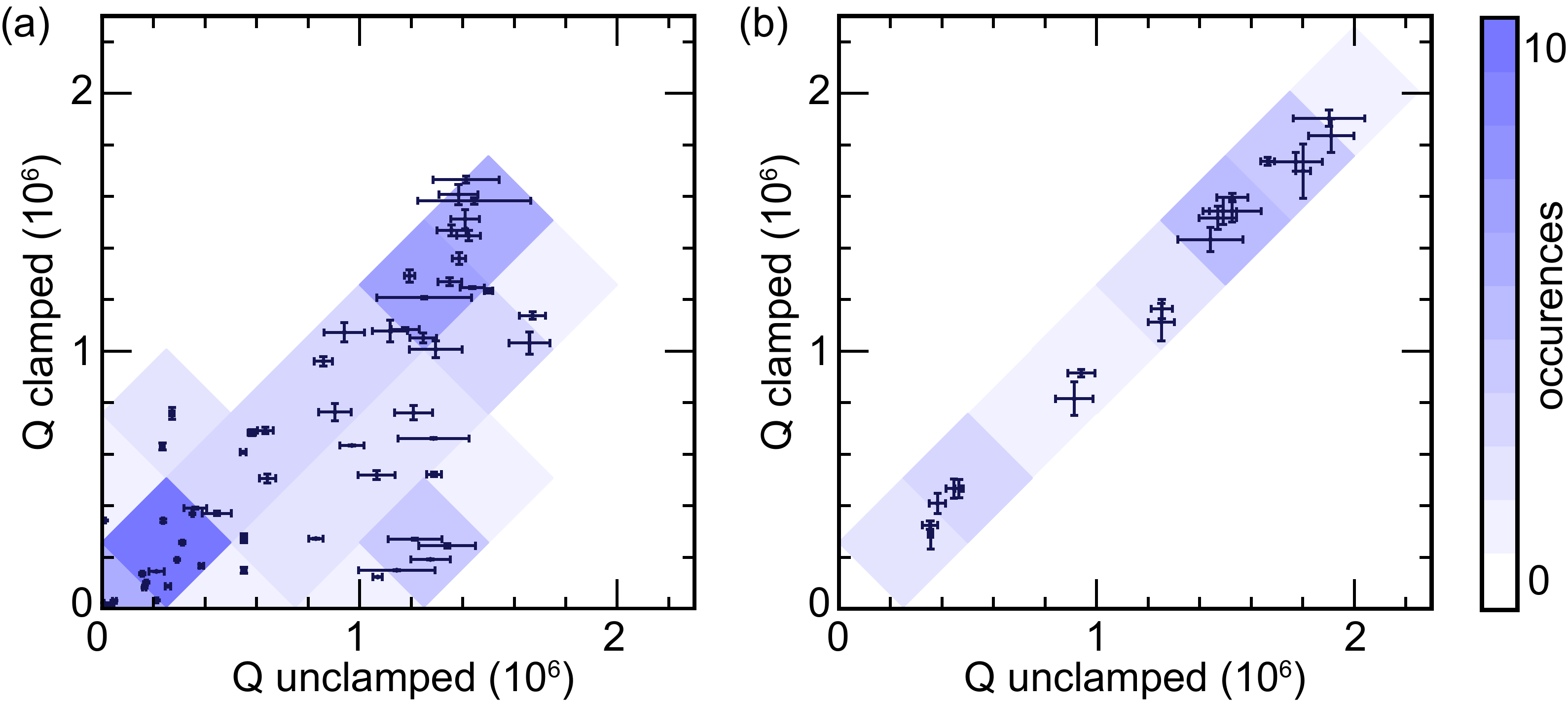}
	\caption{The quality factors for the devices bolted firmly down, compared to the quality factors of the free-standing devices. The 77 mechanical modes are distributed among 7 devices, all three phononic structures (Fig.~\ref{fig:samples}), different membrane sizes, and mode numbers. We distinguish between membrane modes (a) outside and (b) inside phononic bandgaps.}
	\label{fig:overview}
\end{figure}

Finally, as the main result of our work, we present a large compilation of quality factors of the membrane modes, measured on seven different devices, with phononic bridges of all three types shown in Fig.~\ref{fig:samples} and membrane sizes ranging from $\sim 270~\mathrm{\mu m}$ to $\sim 370~\mathrm{\mu m}$.
We separate the in-bandgap modes from the membrane modes outside the bandgap and consider their $Q$-dependence on clamping. 
Fig. \ref{fig:overview} clearly shows that modes inside the bandgap are lined up on the diagonal and are thus unaffected by the clamping of the substrate, while modes outside the bandgap are scattered across the measured landscape.
By measuring a large number of membrane modes, we have thus established the effectivity of the phononic bandgap shield.

\section{Discussion and outlook}
We have shown that the quality factor of the membrane modes can be rendered virtually unaffected by the clamping, and resonant modes of the frame, which indicates that losses due to phonon tunneling into the environment have been eliminated.
Still, not all membrane modes reach the highest quality factors ($\sim 2\times 10^6$).
We attribute this lack of full reproducibility to fabrication imperfections. 
Indeed, we have observed defects and contaminants on the membranes of several devices used for the measurements presented here.
We thus strive to improve fabrication and cleaning protocols.

The highest quality factors are expected to be limited by a membrane-internal dissipation mechanism. 
Assuming constant intrinsic losses it can thus be expected \cite{Yu2012} that higher quality factors can be achieved by altering the dimensions of the membrane resonators. 
For membrane modes with small mode indices the quality factors are proportional to the ratio of the resonator dimensions, $Q\sim L/h$, where $L$ is the side length of the membrane and $h$ is the thickness. 
Thus by increasing the size of the membranes and / or using thinner SiN layers, we can achieve higher $Q$-factors.

\begin{figure}[htb]
	\centering
	\includegraphics[width=\linewidth]{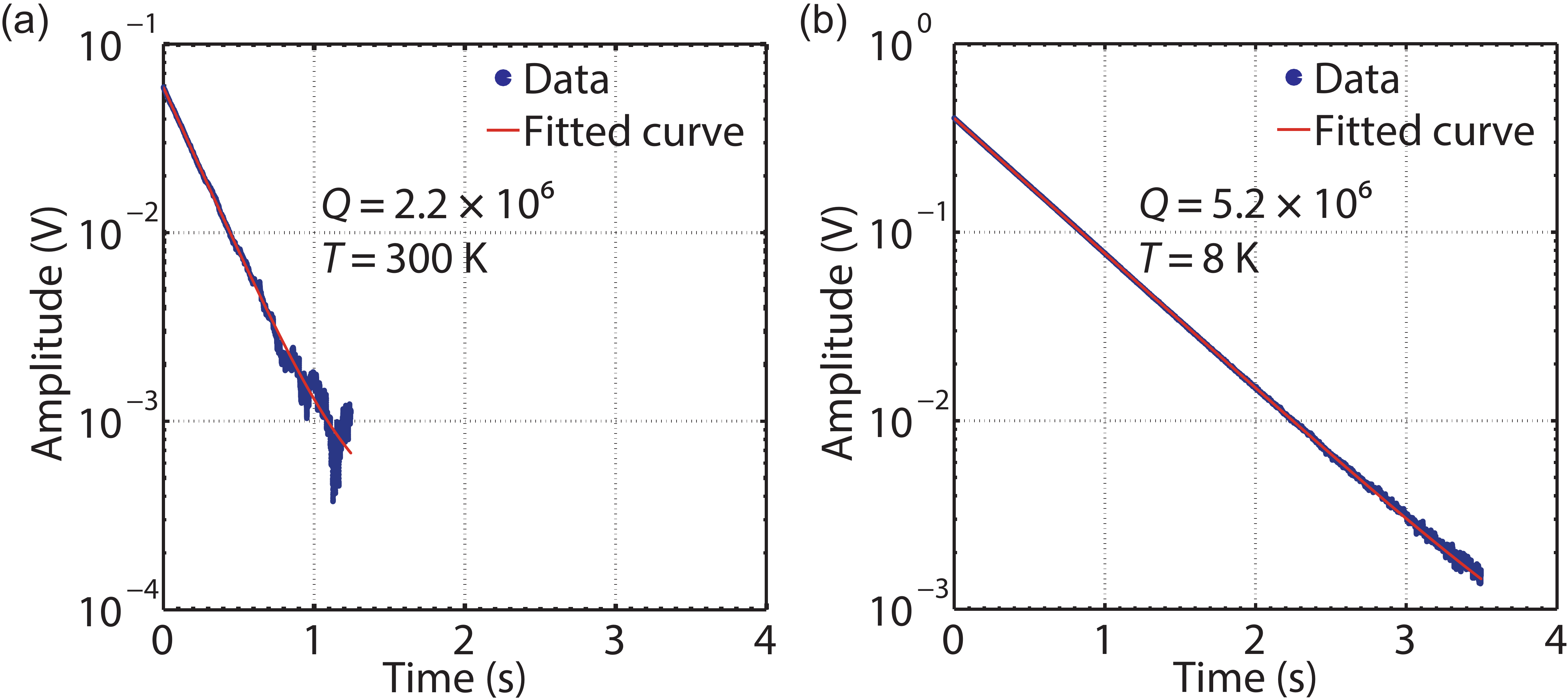}
	\caption{Results of two ringdown measurements used to determine the quality factor of a in-bandgap (2,2) mode for a cross structure device measured at $300~\mathrm{K}$ (a) and $8~\mathrm{K}$ (b). The frequencies are $f = 2.81\times 10^6~\mathrm{Hz}$ and $f = 2.73\times 10^6~\mathrm{Hz}$ at $300~\mathrm{K}$ and $8~\mathrm{K}$, respectively. The vacuum pressure for the room temperature measurement is $\sim 5\times 10^{-7}~\mathrm{mbar}$. The membrane size is $L\sim 285~\mathrm{\mu m}$.}
	\label{fig:cryogenic}
\end{figure}

A strong indication of the dominating role of the internal dissipation mechanism comes from a systematic increase of quality factors when the devices with the highest $Q$'s are cooled to cryogenic temperatures.
In a preliminary study, we have cooled two samples using a $^4$He flow cryostat, providing a high-vacuum below $3\times 10^{-6}~\mathrm{mbar}$, and a cold finger that can be cooled to 8~K.
For this study, we have fabricated a sample holder from oxygen-free, high thermal conductivity (OFHC) copper, which tightly clamps the sample, alongside a spacer and a piezoelectric actuator, in a similar fashion as described before (cf.~Section~4).
Figure~\ref{fig:cryogenic} shows the result obtained on one sample. We have reached a quality factor above $5\times 10^6$ for 4 different in-bandgap modes, distributed among two devices. 
The detailed temperature dependence, the underlying dissipation mechanisms \cite{Enss2005, Southworth2009, Faust2013} and the associated maximum quality factors will be subject of a future study. Furthermore, we believe that the patterning of the silicon substrate will not affect the thermal properties of our devices. While phononic crystals structure can affect the heat flow in materials, the length scales for such periodic structures is typically on the order of tens of nanometers \cite{Maldovan2013}, several order of magnitude below our structures.

Finally, looking back at Fig. \ref{fig:banddiagrams}, it is interesting how different the features of various phononic structures can be.
This raises the question whether it is possible to tailor phononic crystal structures with certain features (e.g. number, position and size of bandgaps).
The complexity of the band diagrams and the many possible choices for the geometry of a unit cell call for an adequate optimization strategy.
Therefore, we have explored the potential of a genetic algorithm, defining the fitness function $F$ to be optimized as the relative size of the bandgap $F\equiv \Delta f / f_c$.
\begin{figure}[htb]
	\centering
	\includegraphics[width=0.4\linewidth]{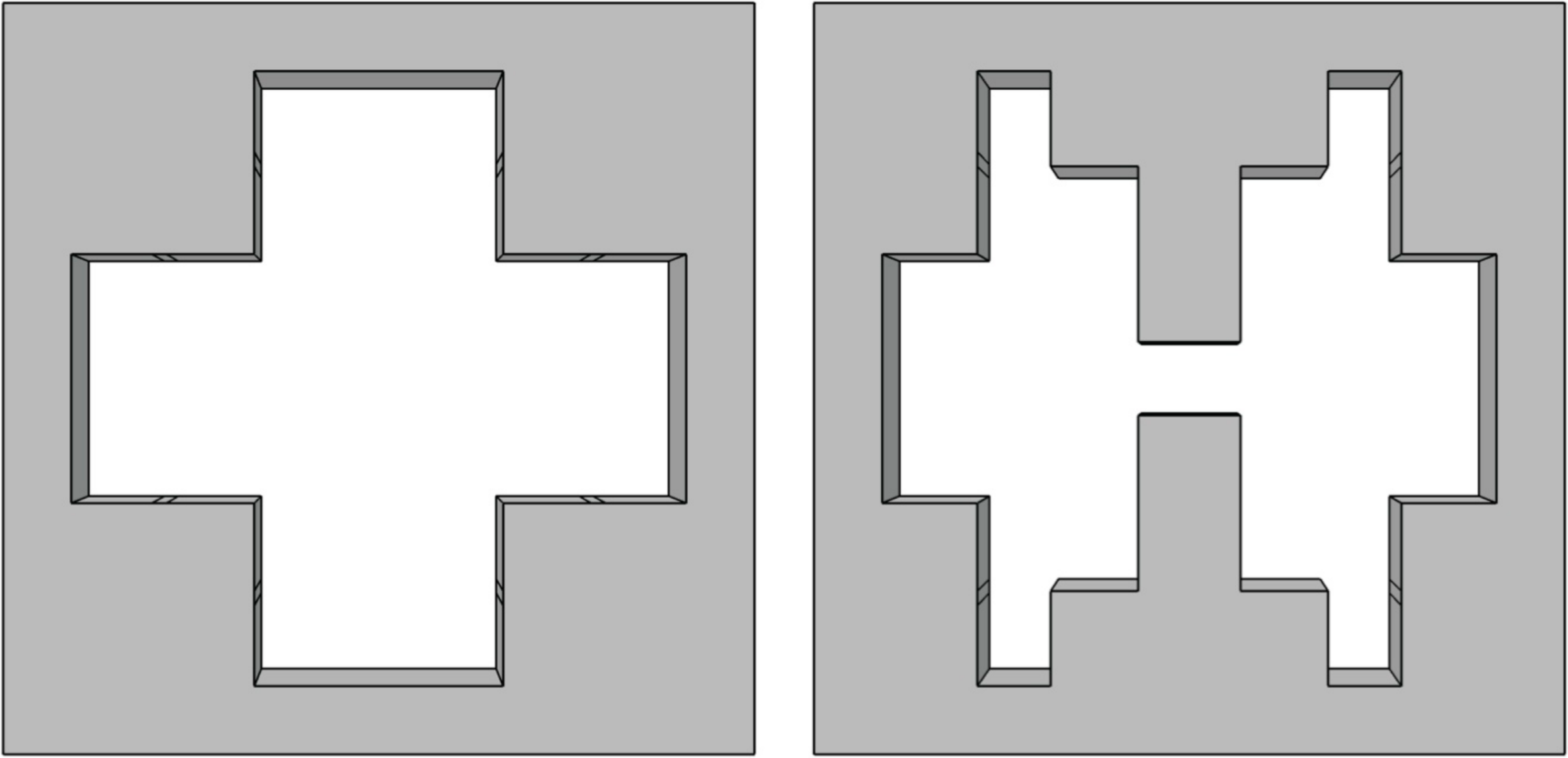}
	\caption{Unit cell geometries found by genetic optimization. The cross structure (left) and the new geometry (right) have bandgap sizes of $\sim 540~\mathrm{kHz}$ and $\sim 170~\mathrm{kHz}$, centered around $\sim 2.6~\mathrm{MHz}$ and $\sim 1.1~\mathrm{MHz}$, respectively.}
	\label{fig:walnut}
\end{figure}
\\

The geometry of the phononic structure's unit cell was parametrized by dividing it into a square grid, which is associated with a binary matrix. 
The value of each matrix element dictates whether in a given grid point there should be silicon or no silicon. 
This approach accommodates the requirements set by our experimental and fabricational constraints. 
In case of a 4-fold symmetric unit cell, the two solutions with the highest fitness found were the cross structure and a new, ``walnut''-like geometry (see Fig. \ref{fig:walnut}). 
Once a suitable geometry has been found, one can employ a different optimization method, such as simulated annealing, to optimize the dimensions of the unit cell for a given geometry.
\section{Conclusion}
In summary, we have here presented a systematic approach in suppressing clamping losses in high-$Q$ silicon nitride resonators, by means of phononic structure shielding. We have demonstrated the suppression through driven response measurements of the silicon structures, as well as a study of the mechanical quality factors of seven devices in two different mounting configurations (firmly clamped frame and free-standing sample). The suppression of clamping losses enables us to tightly clamp our samples to the cryostat cold finger, thus ensuring  good thermalization of the membrane resonators. In that manner, we have reached a quality factor of $5.2\times~10^6$ with a sample tightly bolted to the 8K-cold finger of a cryostat. Our results demonstrate that losses associated with mounting of mechanical resonators can effectively be eliminated by patterning the substrate material.

\section*{Acknowledgments}
This work was supported by the DARPA project QUASAR, the European Union Seventh Framework Program through SIQS (grant no. 600645) and iQOEMS (grant no. 323924), as well as the ERC grants INTERFACE (grant no. 291038). The authors would like to acknowledge valuable discussions with C.~Regal, P.-L.~Yu and T.~P.~Purdy in the initial stage of the project, and their hospitable welcome of Y. T. in Boulder, Colorado, during which the original idea was conceived; as well as Karen Birkelund from DTU Danchip for technical support with microfabrication. S.S acknowledges funding from Villum Foundation's Young Investigator Programme (project no. VKR023125).
\end{document}